\documentclass[12pt,preprint]{aastex}

\slugcomment{ApJS accepted}

\shorttitle{Ambipolar Diffusion in a MHD Code}
\shortauthors{Choi, Kim, \& Wiita}

\begin{document}

\title{An Explicit Scheme for Incorporating Ambipolar Diffusion in a
       Magnetohydrodynamics Code}

\author{Eunwoo Choi\altaffilmark{1}, Jongsoo Kim\altaffilmark{1,2}, and
        Paul J. Wiita\altaffilmark{3,4}}

\altaffiltext{1}{International Center for Astrophysics, Korea Astronomy
                 and Space Science Institute, Daejeon 305-348, Korea;
                 echoi@kasi.re.kr, jskim@kasi.re.kr.}
\altaffiltext{2}{Cavendish Laboratory, University of Cambridge, JJ
                 Thomson Avenue, Cambridge CB3 0HE, UK.}
\altaffiltext{3}{Department of Physics and Astronomy, Georgia State
                 University, P.O. Box 4106, Atlanta, GA 30302-4106;
                 wiita@chara.gsu.edu.}
\altaffiltext{4}{School of Natural Sciences, Institute for Advanced
                 Study, Einstein Drive, Princeton, NJ 08540.}

\begin{abstract}

We describe a method for incorporating ambipolar diffusion
in the strong coupling approximation into a multidimensional
magnetohydrodynamics code based on the total variation diminishing
scheme.
Contributions from ambipolar diffusion terms are included by explicit
finite difference operators in a fully unsplit way, maintaining second
order accuracy.
The divergence-free condition of magnetic fields is exactly ensured at
all times by a flux-interpolated constrained transport scheme.
The super time stepping method is used to accelerate the timestep in
high resolution calculations and/or in strong ambipolar diffusion.
We perform two test problems, the steady-state oblique C-type shocks
and the decay of Alfv\'en waves, confirming the accuracy and robustness
of our numerical approach.
Results from the simulations of the compressible MHD turbulence with
ambipolar diffusion show the flexibility of our method as well as its
ability to follow complex MHD flows in the presence of ambipolar
diffusion.
These simulations show that the dissipation rate of MHD turbulence is
strongly affected by the strength of ambipolar diffusion.

\end{abstract}


\section{Introduction}

In galactic molecular clouds, ambipolar diffusion, which arises in
partially ionized plasmas, is a key ingredient of the mechanism
of star formation \citep[e.g.,][]{mes56,mou76,shu87}.
In the central portions of molecular clouds, the molecular gas is dense
enough that recombination is nearly total, so that very low fractions
($\lesssim10^{-7}$) of the gas remains ionized while the rest of the gas is neutral
\citep{mye85}.
This small residual ionization is usually attributed to cosmic rays, which
can penetrate nearly all clouds.
Ambipolar diffusion causes the relative drift of ions coupled to the
magnetic field and neutrals in the molecular cloud cores and so it
enable the cloud cores to collapse gravitationally.

Star formation assisted by ambipolar diffusion
has been studied extensively in the context of magnetically subcritical
or supercritical models \citep[see e.g.,][]{mou99,des01,bas04,tas07}.
In a current paradigm of star formation, magnetically supported molecular
cloud cores must lose magnetic support through the action of ambipolar
diffusion so that star formation can take place.
Recent works have focused on the role of turbulence in the formation of
protostellar cores \citep[e.g.,][]{nak08}.
Including the effect of turbulence on the mechanism of ambipolar
diffusion can enhance the ambipolar diffusion rate
\citep{fat02,zwe02,hei04}, so that the ambipolar diffusion timescale is
significantly shorter than that estimated for a similar, but quiescent,
medium.
Using three-dimensional numerical simulations, \citet{ois06} and
\citet{li08} have studied the properties of turbulence with ambipolar
diffusion in a two-fluid approximation, while \citet{pad00} have
investigated the heating through ambipolar diffusion in turbulent
molecular clouds using a single-fluid approximation.

Shock waves in molecular clouds spread into steady-state continuous
shocks, or C-type shocks, through ambipolar diffusion.
If the shock speed is slower than the ion Alfv\'en speed but faster than
the neutral sound speed, the ions coupled to magnetic fields drag the
neutrals into the postshock region, producing a continuous structure
\citep{dra80}.
These steady-state C-type shocks can, however, be unstable on a short
enough timescale to be of astrophysical interest.
\citet{war91} showed that if the magnetic field lines are perturbed
slightly and ions collect in the magnetic valleys, the ion-neutral
friction may overcome the magnetic forces in the shock front and derive
an exponentially growing instability.

Numerical treatments of ambipolar diffusion have been commonly derived
from ideal magnetohydrodynamic (MHD) models.
Extensive numerical methods including ambipolar diffusion have been
proposed in the study of the dynamics of partially ionized plasmas within
the frame of single or two fluid models
\citep{tot94,mac95,mac97,smi97,sto97,li06,til08}.
\citet{mac95} have described an explicit method for one-fluid ambipolar
diffusion in the strong coupling limit, while \citet{tot94} has used a
semi-implicit scheme for two-fluid ambipolar diffusion to investigate
instability in C-type shocks.
\citet{til08} have also presented a semi-implicit method for ambipolar
diffusion using a two-fluid approximation.
Implicit schemes for the multifluid treatment of Hall diffusion and
ambipolar diffusion have been suggested by \citet{fal03} and
\citet{osu06,osu07}.

In this work we describe a fully explicit method for incorporating the
single-fluid ambipolar diffusion into a multidimensional MHD code based
on the total variation diminishing scheme.
The divergence-free condition of the magnetic field is ensured by a
flux-interpolated constrained transport scheme, and a super time stepping
method is used in order to considerably accelerate the otherwise
painfully short diffusion-driven time steps.

The organization of this paper is as follows.
In \S 2 the MHD equations are presented along with the approximations we
have made and in \S 3 our numerical methods are described in detail.
Two test problems are presented in \S 4, while MHD turbulence simulations
with significant ambipolar diffusion follow in \S 5.
A summary is given in \S 6.

\section{MHD Equations with Ambipolar Diffusion}

We assume the strong coupling approximation, i.e., that the ion pressure
and momentum are usually negligible in the weakly ionized plasma compared
to those of the neutrals and so the magnetic force on the ions and the
drag force exerted by the neutrals on the ions are almost equal.
In this approximation the plasma can be represented as a single fluid.
This single-fluid approximation turns out to be useful in the formation
of molecular cloud cores through the process of ambipolar diffusion
\citep{shu87}.
To simplify the modeling of ambipolar diffusion here we assume
isothermality with a constant sound speed in the ions and the neutrals
and we ignore gravity.

The isothermal MHD equations including ambipolar diffusion can then be
written as
\begin{equation}
\frac{\partial\rho}{\partial t}
+\mbox{\boldmath$\nabla\cdot$}\left(\rho\mbox{\boldmath$v$}\right) = 0,
\end{equation}
\begin{equation}
\frac{\partial\mbox{\boldmath$v$}}{\partial t}
+\mbox{\boldmath$v\cdot\nabla v$}
+\frac{1}{\rho}\mbox{\boldmath$\nabla$}p
-\frac{1}{\rho}\left(\mbox{\boldmath$\nabla\times B$}\right)
\mbox{\boldmath$\times B$} = 0,
\end{equation}
\begin{equation}
\frac{\partial\mbox{\boldmath$B$}}{\partial t}
-\mbox{\boldmath$\nabla\times$}\left(\mbox{\boldmath$v\times B$}\right)
= \mbox{\boldmath$\nabla\times$}\left\{\left[\frac{1}{\gamma\rho_i\rho}
\left(\mbox{\boldmath$\nabla\times B$}\right)\mbox{\boldmath$\times B$}
\right]\mbox{\boldmath$\times B$}\right\},
\end{equation}
with the additional constraint for the absence of magnetic monopoles,
\begin{equation}
\mbox{\boldmath$\nabla\cdot B$} = 0.
\end{equation}
Here the equation of state is $p=a^2\rho$, where $a$ is an isothermal
sound speed, $\gamma$ is the collisional coupling constant between ions
and neutrals, and $\rho_i$ is the ion density.
The other variables $\rho$, \mbox{\boldmath$v$}, and \mbox{\boldmath$B$}
denote neutral density, neutral velocity, and magnetic field,
respectively.
We renormalize the magnetic field by defining
$\mbox{\boldmath$B$}\equiv\mbox{\boldmath$B$}/\sqrt{4\pi}$ throughout
this paper so that the factor of $4\pi$ does not appear in equations (2)
and (3).

Although the ion density in molecular clouds depends on complicated
physical balance between the cosmic-ray ionization of neutrals and the
recombination of ions and electrons on charged grains, for the purpose
of simplicity we assume that the ion density scales as a power of the
neutral density \citep[e.g.,][]{elm79}
\begin{equation}
\rho_i = \rho_{i0}\left(\frac{\rho}{\rho_0}\right)^\alpha.
\end{equation}
By setting $\alpha=0$ we further simplify by taking the ion density
constant in this work, as these variations will not significantly affect
the problems we treat.
For conditions appropriate to molecular clouds, however, the choice of
$\alpha\sim0.5$ usually would be more realistic, with $\alpha\sim0$
most applicable at very high densities where
grains are the main charge carriers \citep[e.g.,][]{cio98}.
On the other hand, the ion velocity is obtained
from the equation for the relative drift velocity between ions and
neutrals,
\begin{equation}
\mbox{\boldmath$v$}_i = \mbox{\boldmath$v$}+\frac{1}{\gamma\rho_i\rho}
\left(\mbox{\boldmath$\nabla\times B$}\right)\mbox{\boldmath$\times B$}.
\end{equation}
This equation shows that the drag force and the magnetic force on the
ions are balanced and that the ion-neutral drift velocity
$\mbox{\boldmath$v$}_d=\mbox{\boldmath$v$}_i-\mbox{\boldmath$v$}$ is
always perpendicular to the magnetic field.

The basic effect of ambipolar diffusion on the magnetic field can be
expressed in a diffusion coefficient $D$ \citep{shu87} given by
\begin{equation}
D = \tau c_\mathrm{A}^2,
\end{equation}
where $\tau=1/\gamma\rho_i$ is the mean collisional time between ions
and neutrals and $c_\mathrm{A}=B/\sqrt{\rho}$ is the Alfv\'en speed.
Then we can estimate the ambipolar diffusion timescale as
\begin{equation}
t_\mathrm{AD} = \frac{L^2}{D},
\end{equation}
where $L$ is the characteristic length scale of magnetic field.

\section{Numerical Methods}

\subsection{Source Term Integration}

The numerical scheme for solving the ideal MHD equations is described
in previous works \citep{ryu95,ryu98,kim99}.
This method is based on the total variation diminishing (TVD) scheme
\citep{har83} which is an explicit Eulerian upwind scheme with a
second-order accuracy in space and time.
In the strong coupling approximation we can separate the ion density and
velocity from the neutral density and velocity so that we can basically
use the MHD TVD code to compute the evolution of the neutral density and
velocity using mass and momentum conservation equations.

We now describe how to incorporate ambipolar diffusion terms into the
induction equation.
The induction equation can be rewritten in component form as
\begin{equation}
\frac{\partial B_x}{\partial t}
+\frac{\partial}{\partial y}\left(B_x v_y-B_y v_x\right)
-\frac{\partial}{\partial z}\left(B_z v_x-B_x v_z\right)
= \frac{\partial S_z}{\partial y}-\frac{\partial S_y}{\partial z},
\end{equation}
\begin{equation}
\frac{\partial B_y}{\partial t}
+\frac{\partial}{\partial z}\left(B_y v_z-B_z v_y\right)
-\frac{\partial}{\partial x}\left(B_x v_y-B_y v_x\right)
= \frac{\partial S_x}{\partial z}-\frac{\partial S_z}{\partial x},
\end{equation}
\begin{equation}
\frac{\partial B_z}{\partial t}
+\frac{\partial}{\partial x}\left(B_z v_x-B_x v_z\right)
-\frac{\partial}{\partial y}\left(B_y v_z-B_z v_y\right)
= \frac{\partial S_y}{\partial x}-\frac{\partial S_x}{\partial y}.
\end{equation}
Here the source term components are given by
\begin{equation}
S_x = \frac{1}{\gamma\rho_i\rho}\left[\left(\frac{\partial B_y}
{\partial x}-\frac{\partial B_x}{\partial y}\right)B_x B_z
+\left(\frac{\partial B_y}{\partial z}-\frac{\partial B_z}
{\partial y}\right)\left(B_z^2+B_y^2\right)+\left(\frac{\partial B_x}
{\partial z}-\frac{\partial B_z}{\partial x}\right)B_x B_y\right],
\end{equation}
\begin{equation}
S_y = \frac{1}{\gamma\rho_i\rho}\left[\left(\frac{\partial B_z}
{\partial y}-\frac{\partial B_y}{\partial z}\right)B_y B_x
+\left(\frac{\partial B_z}{\partial x}-\frac{\partial B_x}
{\partial z}\right)\left(B_x^2+B_z^2\right)+\left(\frac{\partial B_y}
{\partial x}-\frac{\partial B_x}{\partial y}\right)B_y B_z\right],
\end{equation}
\begin{equation}
S_z = \frac{1}{\gamma\rho_i\rho}\left[\left(\frac{\partial B_x}
{\partial z}-\frac{\partial B_z}{\partial x}\right)B_z B_y
+\left(\frac{\partial B_x}{\partial y}-\frac{\partial B_y}
{\partial x}\right)\left(B_y^2+B_x^2\right)+\left(\frac{\partial B_z}
{\partial y}-\frac{\partial B_y}{\partial z}\right)B_z B_x\right].
\end{equation}
Note that the source terms on the right-hand side of equations (9) to (11)
have the same divergence form as the flux components on the left-hand side.

Standard second-order finite difference operators are applied to the
explicit discretization of the source components.
Here we define the source components at grid centers, $S_{x,i,j,k}$,
$S_{y,i,j,k}$, and $S_{z,i,j,k}$, while the $n$-th components of the
TVD flux vectors in each direction, $\bar{f}_{x,i+1/2,j,k}^{(n)}$,
$\bar{f}_{y,i,j+1/2,k}^{(n)}$, and $\bar{f}_{z,i,j,k+1/2}^{(n)}$, are
defined at face centers.
The first four components of the TVD flux vectors, $\bar{f}^{(1)}$
through $\bar{f}^{(4)}$, are the upwind fluxes associated with the
transport of mass and momentum, i.e., mass and momentum advection fluxes,
and the last three components of the TVD flux vectors, $\bar{f}^{(5)}$
through $\bar{f}^{(7)}$, represent the components of the electric field.
While keeping second-order accuracy, the source components can then be
included in the TVD flux components as follows
\begin{equation}
f_{x,i+1/2,j,k}^{(6)} = \bar{f}_{x,i+1/2,j,k}^{(6)}
+\frac{1}{2}\left(S_{z,i,j,k}+S_{z,i+1,j,k}\right),
\end{equation}
\begin{equation}
f_{x,i+1/2,j,k}^{(7)} = \bar{f}_{x,i+1/2,j,k}^{(7)}
-\frac{1}{2}\left(S_{y,i,j,k}+S_{y,i+1,j,k}\right),
\end{equation}
\begin{equation}
f_{y,i,j+1/2,k}^{(7)} = \bar{f}_{y,i,j+1/2,k}^{(7)}
+\frac{1}{2}\left(S_{x,i,j,k}+S_{x,i,j+1,k}\right),
\end{equation}
\begin{equation}
f_{y,i,j+1/2,k}^{(5)} = \bar{f}_{y,i,j+1/2,k}^{(5)}
-\frac{1}{2}\left(S_{z,i,j,k}+S_{z,i,j+1,k}\right),
\end{equation}
\begin{equation}
f_{z,i,j,k+1/2}^{(5)} = \bar{f}_{z,i,j,k+1/2}^{(5)}
+\frac{1}{2}\left(S_{y,i,j,k}+S_{y,i,j,k+1}\right),
\end{equation}
\begin{equation}
f_{z,i,j,k+1/2}^{(6)} = \bar{f}_{z,i,j,k+1/2}^{(6)}
-\frac{1}{2}\left(S_{x,i,j,k}+S_{x,i,j,k+1}\right).
\end{equation}
Since the TVD scheme has second-order accuracy, the above second-order
interpolation of the source components should be adequate.
The contributions of the source components are added in a fully unsplit
way after all the TVD flux components are updated through the TVD step.
These total advective fluxes at face centers, $f_{x,i+1/2,j,k}^{(n)}$,
$f_{y,i,j+1/2,k}^{(n)}$, and $f_{z,i,j,k+1/2}^{(n)}$, are used to enforce
$\mbox{\boldmath$\nabla\cdot B$}=0$ as described in the following
subsection as well as to update the magnetic field components to the next
time step.

\subsection{Divergence-Free Condition}

Analytically the divergence-free condition
$\mbox{\boldmath$\nabla\cdot B$}=0$ is maintained if the condition holds
for the initial magnetic field, but numerically the divergence of the
magnetic field will not be exactly zero due to numerical discretization
and dimensional splitting.
Several schemes to maintain $\mbox{\boldmath$\nabla\cdot B$}=0$ constraint
have been suggested and used in numerical MHD \citep[see][]{tot00}.
\citet{eva88} suggested the constrained transport (CT) scheme which used
a specific finite difference discretization on a staggered mesh to satisfy
the divergence-free constraint.
The flux-interpolated CT schemes \citep{ryu98,bal99} have introduced a new
staggered magnetic field variable which is updated by simple finite
differences using the interpolated fluxes.
We have found the flux-interpolated CT scheme to be effective for
incorporating ambipolar diffusion and have followed the approach suggested
by \citet{bal99}.

Using the total fluxes at face centers, equations (15) to (20), the
advective fluxes at grid edges are reconstructed with second-order
accuracy as follows
\begin{equation}
\Omega_{x,i,j+1/2,k+1/2} = \frac{1}{4}
\left(f_{z,i,j,k+1/2}^{(6)}+f_{z,i,j+1,k+1/2}^{(6)}
-f_{y,i,j+1/2,k}^{(7)}-f_{y,i,j+1/2,k+1}^{(7)}\right),
\end{equation}
\begin{equation}
\Omega_{y,i+1/2,j,k+1/2} = \frac{1}{4}
\left(f_{x,i+1/2,j,k}^{(7)}+f_{x,i+1/2,j,k+1}^{(7)}
-f_{z,i,j,k+1/2}^{(5)}-f_{z,i+1,j,k+1/2}^{(5)}\right),
\end{equation}
\begin{equation}
\Omega_{z,i+1/2,j+1/2,k} = \frac{1}{4}
\left(f_{y,i,j+1/2,k}^{(5)}+f_{y,i+1,j+1/2,k}^{(5)}
-f_{x,i+1/2,j,k}^{(6)}-f_{x,i+1/2,j+1,k}^{(6)}\right).
\end{equation}
Then the magnetic field components at face centers are updated as
\begin{eqnarray}
b_{x,i+1/2,j,k}^{n+1} = b_{x,i+1/2,j,k}^n \nonumber
&-&\frac{\Delta t}{\Delta y}
\left(\Omega_{z,i+1/2,j+1/2,k}-\Omega_{z,i+1/2,j-1/2,k}\right) \\
&+&\frac{\Delta t}{\Delta z}
\left(\Omega_{y,i+1/2,j,k+1/2}-\Omega_{y,i+1/2,j,k-1/2}\right),\\
b_{y,i,j+1/2,k}^{n+1} = b_{y,i,j+1/2,k}^n \nonumber
&-&\frac{\Delta t}{\Delta z}
\left(\Omega_{x,i,j+1/2,k+1/2}-\Omega_{x,i,j+1/2,k-1/2}\right) \\
&+&\frac{\Delta t}{\Delta x}
\left(\Omega_{z,i+1/2,j+1/2,k}-\Omega_{z,i-1/2,j+1/2,k}\right),\\
b_{z,i,j,k+1/2}^{n+1} = b_{z,i,j,k+1/2}^n \nonumber
&-&\frac{\Delta t}{\Delta x}
\left(\Omega_{y,i+1/2,j,k+1/2}-\Omega_{y,i-1/2,j,k+1/2}\right) \\
&+&\frac{\Delta t}{\Delta y}
\left(\Omega_{x,i,j+1/2,k+1/2}-\Omega_{x,i,j-1/2,k+1/2}\right).
\end{eqnarray}
It is straightforward to show that
$\mbox{\boldmath$\nabla\cdot b$}^{n+1}=\mbox{\boldmath$\nabla\cdot b$}^n=0$
if the numerical divergence of \mbox{\boldmath$b$} is initially zero.

In the TVD scheme for MHD, all fluid quantities are defined at grid
centers.
Thus the magnetic field components at grid centers are interpolated as
\begin{equation}
B_{x,i,j,k} = \frac{1}{2}\left(b_{x,i+1/2,j,k}+b_{x,i-1/2,j,k}\right),
\end{equation}
\begin{equation}
B_{y,i,j,k} = \frac{1}{2}\left(b_{y,i,j+1/2,k}+b_{y,i,j-1/2,k}\right),
\end{equation}
\begin{equation}
B_{z,i,j,k} = \frac{1}{2}\left(b_{z,i,j,k+1/2}+b_{z,i,j,k-1/2}\right).
\end{equation}
Note that the above arithmetic interpolation will be sufficient to
maintain second-order accuracy.

\subsection{Super Time Stepping}

The time step for ambipolar diffusion is proportional to the square
of the grid size in the single fluid approximation, so the explicit
treatment of ambipolar diffusion terms leads to very small time steps
\citep{mac95}.
In the two fluid approximation that treats ions and neutrals separately,
including the ion momentum equation severely limits the time steps via a
very restrictive stability criterion.
To resolve this problem, previous papers have proposed different
solutions.
For instance, \citet{li06} proposed a ``heavy-ion'' approximation to
speed up the time steps, while \citet{nak08} set a density threshold
below which the ambipolar diffusion rate is set to zero to avoid this
problem.
In this work we adopt the ``super time stepping'' approach
\citep{ale96} to increase the effective time interval and allow much
faster computations for ambipolar diffusion.
\citet{osu06,osu07} also used this strategy in their multifluid MHD
models.

The super time stepping technique considerably accelerates the explicit
schemes for parabolic problems \citep{ale96}.  The key advantage of this
approach is that it demands stability over large compound time steps,
rather than over each of the constituent substeps.
In this method, the state vector is evolved over a super time step,
\begin{equation}
\Delta t_\mathrm{STS} = \Delta t_\mathrm{AD}\frac{n}{2\sqrt{\nu}}\left[
\frac{\left(1+\sqrt{\nu}\right)^{2n}-\left(1-\sqrt{\nu}\right)^{2n}}
{\left(1+\sqrt{\nu}\right)^{2n}+\left(1-\sqrt{\nu}\right)^{2n}}\right],
\end{equation}
where $\Delta t_\mathrm{AD}$ is the nominal ambipolar diffusion timestep,
$n$ is the number of substeps, and $\nu$ is a fuzzy factor ($0<\nu<1$).
This super time step is defined as
$\Delta t_\mathrm{STS}=\sum_{j=1}^n\Delta\tau_j$, and consists of the
substeps, $\Delta\tau_j$, which are given by
\begin{equation}
\Delta\tau_j = \Delta t_\mathrm{AD}\left[\left(\nu-1\right)
\cos\left(\frac{2j-1}{n}\frac{\pi}{2}\right)+\nu+1\right]^{-1}.
\end{equation}
It has been proven that
$\Delta t_\mathrm{STS}\rightarrow n^2\Delta t_\mathrm{AD}$ as $\nu\rightarrow0$
so that the super time step approach is asymptotically $n$ times faster
than the standard explicit scheme \citep{ale96}.
However, the $\nu$ parameter must be properly chosen for each problem in
order to achieve optimality and stability of performance.
For diffusion-dominated problems, $\Delta t_\mathrm{AD}$ drops below the
typical Courant time step, $\Delta t_\mathrm{Cour}$
(i.e., $\Delta t_\mathrm{AD}<\Delta t_\mathrm{STS}\leq\Delta t_\mathrm{Cour}$).
Thus, if $\Delta t_\mathrm{STS}$ is taken to be the Courant time step,
the super time stepping method requires roughly
$\sqrt{\Delta t_\mathrm{Cour}/\Delta t_\mathrm{AD}}$ substeps.

In addition to allowing larger effective time steps, the super time
stepping approach offers relatively simple implementation since it is a
first order method.
We have successfully applied this approach to the linear and nonlinear
ambipolar diffusion problems presented in the following sections.
Our numerical results confirm the efficiency and accuracy of the super time
stepping approach, as previously implemented in other approximations.

\section{Test Problems}

\subsection{Oblique C-type Shocks}

We compute the structure of oblique C-type shocks in order to test the
numerical methods described in the previous section.
The steady state structure of a C-type shock is characterized
conveniently by the shock length scale defined as
\begin{equation}
L_s = \tau c_\mathrm{A}.
\end{equation}
Following \citet{mac95}, the steady state equations are solved by setting
$\partial_t=\partial_y=\partial_z=0$ in equations (1) to (3), which then
can be reduced to a single ordinary differential equation for $\rho$.
The steady state solution can be obtained through the numerical
integration of the ordinary differential equation.
In a separate code we numerically integrate the ordinary differential
equation using the fourth-order Runge-Kutta method.
This solution is specified by the three parameters, the sonic Mach
number $M$, the Alfv\'en Mach number $M_\mathrm{A}$, and the angle
$\theta$ between the shock normal and the magnetic field.

To generate C-type shocks we set up a two-dimensional shock heating
problem.
Initially a gas with a neutral density $\rho$ propagates with a velocity
\mbox{\boldmath$v$} against a reflecting wall placed at $x=0$ in the
uniform magnetic field \mbox{\boldmath$B$} that lies at an angle $\theta$
to the $x$-axis.
As the gas hits the reflecting wall, both the fluid and the magnetic
field are compressed, the gas is heated and a reverse shock is produced.
The ion-neutral friction drags the neutral gas into the postshock region,
and finally the steady-state C-type shock is built up, yielding the
appropriate continuous transition.
The parameters we chose for this problem are $a=0.1$, $\gamma=1$,
$\rho_i=10^{-5}$, $\rho=1$, and
$\mbox{\boldmath$B$}=B_{0x}\mbox{\boldmath$\hat{x}$}+B_{0y}\mbox{\boldmath$\hat{y}$}$
with $B_{0x}=B_{0y}=1/\sqrt{2}$.
This problem has been set up with two inflow velocities,
$\mbox{\boldmath$v$}=-4.45\mbox{\boldmath$\hat{x}$}$ and
$-9.47\mbox{\boldmath$\hat{x}$}$, which correspond to the shock velocities
$v_s=5$ and $10$ for our chosen parameters.
This gives $M=v_s/a=50$ and $100$, $M_\mathrm{A}=v_s/c_\mathrm{A}=5$ and
$10$, and $\theta=\tan^{-1}(B_{0y}/B_{0x})=\pi/4$.
The computations have been done in a two-dimensional box of $x=y=[0,L]$
with $L=20L_s$ using $128^2$ cells.
Outflow boundary conditions are used except for a reflecting boundary
imposed at $x=0$.

\begin{figure}
\begin{center}
\includegraphics[scale=0.8]{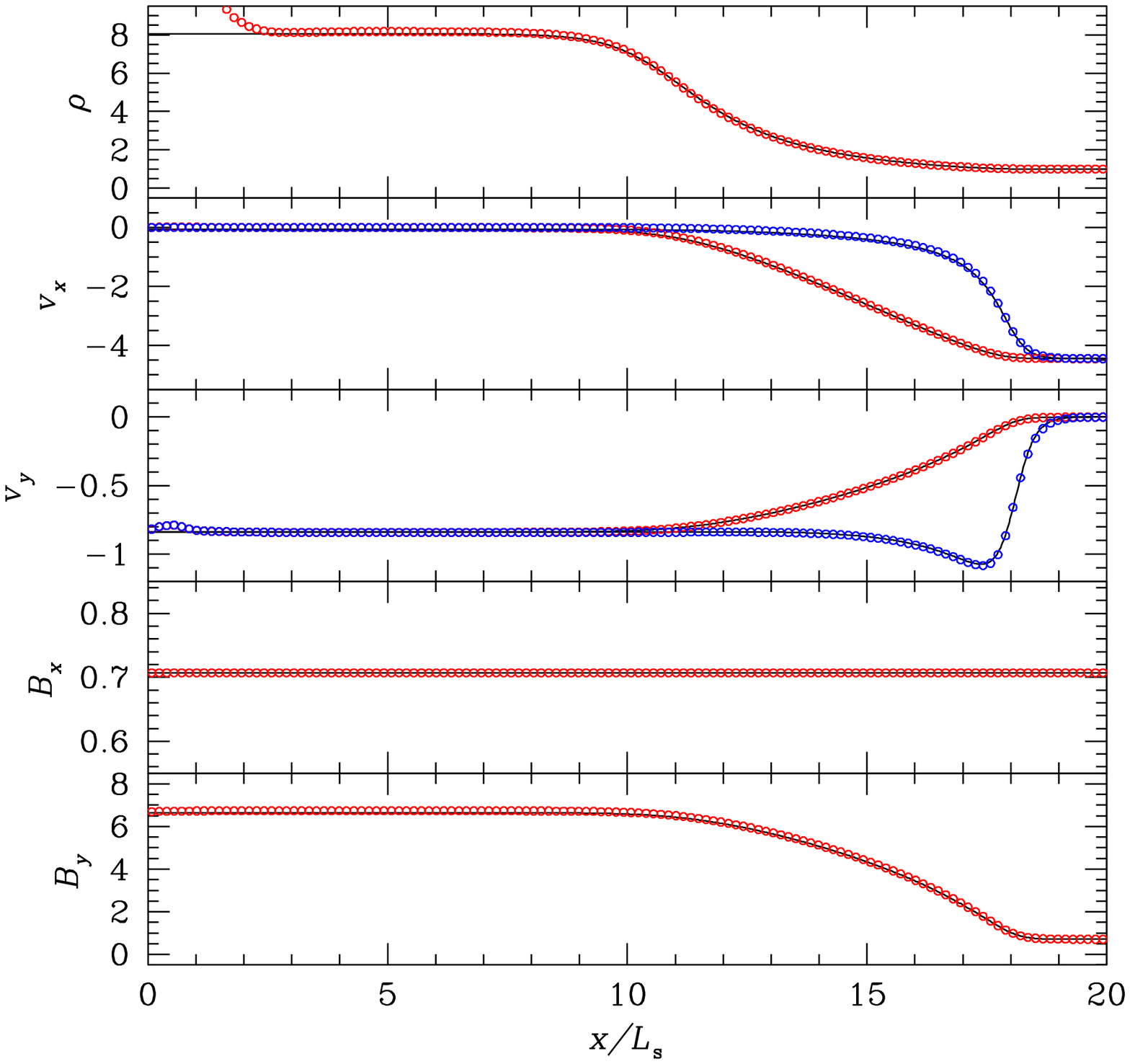}
\end{center}
\caption{(a) Structure of an oblique C-type shock with $M=50$,
$M_\mathrm{A}=5$ and $\theta=\pi/4$.
Profiles of neutral density, neutral (red) and ion (blue) velocity
components, and magnetic field components from the numerical calculation
are marked with open circles.
The solid lines represent the analytic solution for the steady-state
C-type shock.
The calculation has been done in a square box with size $L=20L_s$ using
$128^2$ cells.
(b) Same as in (a) except for $M=100$, $M_\mathrm{A}=10$, and
$\theta=\pi/4$.}
\label{fig1}
\end{figure}

The parameters for the run shown in Figure \ref{fig1}(a) are $M=50$,
$M_\mathrm{A}=5$, and $\theta=\pi/4$ and those for the run shown in
Figure \ref{fig1}(b) are $M=100$, $M_\mathrm{A}=10$, $\theta=\pi/4$.
In Figure \ref{fig1}, the structure of neutral density,
neutral (red) and ion (blue) velocity components, and magnetic field
components from numerical calculations are marked with open circles and
compared to analytic solutions plotted with solid lines.
Structures are measured along the $x$-direction before the shock reaches
the outer boundary.
The spike in the neutral density seen in a few cells
near the reflecting wall is the overheating phenomenon.
This is purely a numerical artifact that most finite
difference schemes applied to the shock heating problem intrinsically
possess since they cannot compute the jump condition across strong shocks
within a single cell.
In all the flow variables, the structure of the C-type shock clearly
forms.
Figure \ref{fig1} shows the excellent agreement between the
numerical solutions and the analytic solutions for the steady-state
C-type shocks, demonstrating the accuracy of our numerical methods.

The accuracy of numerical solutions depends on the number of cells
spanned by the box size $L$.
So we have run the case of the test in Figure \ref{fig1}(a) with
different numerical resolutions to check the convergence properties.
Except for the resolutions the initial conditions are identical to those
used in the test in Figure \ref{fig1}(a).
We have computed the mean errors for neutral density defined by
$\bar{E}(\rho)=\sum_{i,j}|\rho_{i,j}^n-\rho_{i,j}^a|/\sum_{i,j}|\rho_{i,j}^a|$,
where the superscript $n$ represents numerical solution and the
superscript $a$ represents analytic solution.
The resolutions of $16^2$, $32^2$, $64^2$, $128^2$, and $256^2$ cells
give the mean errors of $1.33$, $1.29$, $1.28$, $1.27$, and $1.27$,
respectively.
As expected, the mean errors for neutral density converge as the
numerical resolution increases.
In this convergence test we also see that there are clear trends toward
convergence in mean errors for velocity and magnetic field, and that a
$128^2$ grid is sufficient to treat this problem.

\subsection{Decay of Alfv\'en Waves}

The propagation of Alfv\'en waves in a weakly ionized plasma provides
an effective tool for testing the dynamics of ambipolar diffusion.
\citet{kul69} first showed that ambipolar diffusion can prevent the
propagation of Alfv\'en waves in a partially ionized medium.
In the strong coupling approximation \citet{bal96} gives an
explicit quadratic dispersion relation for Alfv\'en waves,
\begin{equation}
\omega^2+i\frac{c_\mathrm{A}^2k^2}{\gamma\rho_i}\omega
-c_\mathrm{A}^2k^2 = 0,
\end{equation}
where $\omega=\omega_R+i\omega_I$ is the complex angular frequency of
the wave and $k$ is a real wavenumber.
It is clear in the above equation that the Alfv\'en waves always
propagate when $k<2\gamma\rho_i/c_\mathrm{A}$ (i.e., $\omega_R\neq0$).
In order to test the propagation of Alfv\'en waves in the strong
coupling limit, we have followed the evolution of a standing wave in
numerical calculations and compared the oscillation frequency and decay
rate of the wave to the analytic results.
Damped oscillations of standing waves have been long studied
\citep[e.g.,][]{mor86} and the time-dependence of the first normal mode
is described by
\begin{equation}
h\left(t\right) = h_0\left|\sin\left(\omega_R t\right)\right|
e^{\omega_I t},
\end{equation}
where $h_0$ is the initial amplitude of the wave.

In our test of the decay of Alfv\'en waves, we have used a standing
wave formed along the diagonal on $x-y$ plane with initial velocity
\begin{equation}
\mbox{\boldmath$v$} = v_\mathrm{amp}c_\mathrm{A}
\sin\left(k_x x+k_y y\right)\mbox{\boldmath$\hat{z}$}.
\end{equation}
The background density and magnetic field have been set to be uniform
with $\rho=1$ and $\mbox{\boldmath$B$}=B_0\mbox{\boldmath$\hat{x}$}$
with $B_0=1$.
This gives the characteristic Alfv\'en speed
$c_\mathrm{A}=B/\sqrt{2\rho}=0.7071$.
Here the initial peak amplitude has been set to $v_\mathrm{amp}=0.1$
and the wavenumbers have been set to $k_x=k_y=2\pi/L$ so the total
wavenumber is $k=\sqrt{k_x^2+k_y^2}$.
We choose collisional coupling constants, $\gamma=100$, $500$, and
$1000$ with $a=1$ and $\rho_i=0.1$ in this test.
The calculations have been done in a computational box of
$x=y=z=[0,L]$ with $L=1$ using $128^3$ cells.
Boundary conditions are periodic in the $x$- and $y$-directions and
outflow in the $z$-direction.

Figure \ref{fig2} shows the time evolution of the (spatially) root mean
square magnetic field in the $z$-direction, $<\delta B_z^2>^{1/2}$, for
three different collisional coupling constants $\gamma=1000$ (top),
$500$ (middle), and $100$ (bottom) in the test of the decay of Alfv\'en
waves.
Our numerical results are marked with open circles while the theoretical
predictions from equation (34) are plotted as solid lines.
The oscillation frequencies and decay rates from these numerical
calculations fit very well to those from the theoretical predictions,
confirming that our numerical methods are accurate.
Based on the time evolution of the standing wave in Figure \ref{fig2},
the propagation of Alfv\'en waves is significantly suppressed with
decreasing collisional coupling constant $\gamma$.

By fitting the time evolution of $<\delta B_z^2>^{1/2}$ to theoretical
curves from equation (34) we find the numerical data for the complex
angular frequency $\omega$, whose real and imaginary parts correspond to
the oscillation frequency and decay rate of the standing wave,
respectively.
We repeated the calculation of Figure \ref{fig2} for $\gamma=100$, for
eight wavenumbers ranging from $2\pi/16$ to $6\sqrt{2}\pi$, collecting
the data for the oscillation frequencies $\omega_R$ and the decay rates
$\omega_I$.
In Figure \ref{fig3} we show those ``experimental'' results together
with the analytic solution of equation (33).
Oscillation frequencies (red) and decay rates (blue) found from the
numerical experiments are represented with filled circles and the
analytic solutions are drawn as solid lines.
In Figure \ref{fig3} the very good agreement between numerical data and
theoretical predictions for different wavenumbers shows the accuracy and
flexibility of our numerical methods.

\section{MHD Turbulence Simulations}

In this section we present, as a first practical problem using this code,
simulations of the compressible MHD turbulence in the presence of
ambipolar diffusion.
A simulation of turbulent ambipolar diffusion was studied by
\citet{pad00} in the strong coupling approximation and by \citet{ois06}
and \citet{li08} in the two-fluid approximation.
We have performed this simulation to confirm the validation of our
numerical methods for including ambipolar diffusion as well as to
investigate the role of ambipolar diffusion in the dissipation of
compressible MHD turbulence.

To characterize the MHD simulations, we define the strength of the
magnetic field in terms of the parameter
\begin{equation}
\beta \equiv \frac{a^2}{c_\mathrm{A}^2}.
\end{equation}
Note that our definition of $\beta$ differs by a factor of 2 from the
usual plasma $\beta$, the ratio of gas pressure to magnetic pressure.
For MHD turbulent flows with the root mean square velocity
$v_\mathrm{rms}$, the importance of magnetic fields on the dynamics of
gas is characterized by the Alfv\'en Mach number
$M_\mathrm{A}=v_\mathrm{rms}/c_\mathrm{A}=\sqrt{\beta}M$, where the
sonic Mach number $M$ is given by $M=v_\mathrm{rms}/a$.
The effect of ambipolar diffusion on Alfv\'enic turbulence of scale $L$
is measured through the ambipolar diffusion Reynolds number \citep{bal96}
defined by
\begin{equation}
R_\mathrm{AD} \equiv \frac{v_\mathrm{rms}L}{\tau c_\mathrm{A}^2}.
\end{equation}
A sufficiently large ambipolar diffusion Reynolds number implies that
the importance of ambipolar diffusion to the turbulent flow on the scale
$L$ becomes vanishingly small.
The ambipolar diffusion length scale $l_\mathrm{AD}$ is then defined as
a characteristic length scale at which the ambipolar diffusion Reynolds
number becomes unity, i.e.,
$l_\mathrm{AD}=\tau c_\mathrm{A}^2/v_\mathrm{rms}$.

We consider two types of MHD turbulence models driven according to the
method described in \citet{sto98}.
One is turbulence decaying from saturated initial velocity perturbations,
and the other is forced turbulence in which velocity perturbations are
added at constant time intervals.
In both decaying and forced turbulence, the velocity perturbations
$\delta v$ are generated from a Gaussian random field with a power
spectrum
\begin{equation}
P\left(k\right) \equiv \delta v^2\left(k\right)
\propto k^6\exp\left(-8k/k_p\right),
\end{equation}
where the power spectrum peaks at $k_p=4(2\pi/L)$.
The velocity perturbations are subject to the constraints that
$\mbox{\boldmath$\nabla\cdot$}\delta\mbox{\boldmath$v$}=0$ and no net
momentum is added by the velocity perturbations,
$<\rho\delta\mbox{\boldmath$v$}>=0$.
The perturbations are normalized so that the initial kinetic energy
$\delta E_K=50$ for decaying turbulence and a constant kinetic energy
input rate $\dot{E}_K=500$ is injected at regular time intervals
$\delta t=0.001$ for forced turbulence.

The simulation parameters for both decaying models and forced models are
summarized in Table \ref{tab1}.
According to the values of $\gamma$ varying from $\infty$ to $100$, we
denote decaying models as D1 to D3 and forced models as F1 to F3.
In Table \ref{tab1}, the flow time is defined as $t_f=L/v_\mathrm{rms}$,
and $t_\mathrm{end}$ is the simulation end time.
The model simulations have been set up with uniform neutral density
$\rho=1$ and uniform magnetic field 
$\mbox{\boldmath$B$}=B_0\mbox{\boldmath$\hat{x}$}$ with $B_0=1$.
The isothermal sound speed and ion density are assumed to be constant
with $a=1$ and $\rho_i=0.01$.
The simulations have been done in a periodic box of $x=y=z=[0,L]$ with
$L=1$ using $128^3$ cells.

In Figures \ref{fig4}(a) and (b) we present images of the logarithms
of the neutral density for models D1 (top) and D3 (bottom) at $t=1t_f$
and for models F1 (top) and F3 (bottom) at $t=1t_f$, respectively.
The images are slices through the $x-y$ plane at $z=0.5$.
The overall density features for both decaying turbulence and forced
turbulence models are roughly the same, even though the turbulence
driving pattern is different.
Small-scale knots and filaments are produced in the absence of ambipolar
diffusion (models D1 and F1).
By introducing strong ambipolar diffusion (i.e., reducing the
collisional coupling constant to $10$), the density structures diffuse
out, resulting in larger, smoother density structures, as shown in
models D3 and F3.
The results for models D2 and F2 are intermediate and are not shown.
We note that the ranges of observed densities are smaller in the cases
where ambipolar diffusion is strong, regardless of whether the turbulence
is decaying or forced.

Figure \ref{fig5} shows the time evolution of the total energy,
defined as the sum of the kinetic energy and the energy of the perturbed
magnetic field, $E_\mathrm{tot}=E_K+E_B$, where
\begin{equation}
E_K = \frac{1}{2}\int_V\rho v^2dV,
\end{equation}
and
\begin{equation}
E_B = \frac{1}{2}\int_V\left(B^2-B_0^2\right)dV,
\end{equation}
with $dV=dxdydz$.
The evolution of the total energy for the decaying models, D1 (black),
D2 (blue), and D3 (red) are plotted in Figure \ref{fig5}(a).
After initial plateau phases all these models lose their initial energies
rapidly, decaying nearly as
power-laws with time (with indices between $1.1$ to $1.3$).
The rate of the turbulent energy decay significantly increases with
decreases in the collisional coupling constant, and thus we see that the
turbulent decay rate can be strongly affected by differences in ambipolar
diffusion.  The somewhat more rapid decline of the total energy
found for model D1 than for a similar simulation by \citet{sto98} can be understood
in terms of the slightly lower resolution and higher initial
energy in our simulation.
In Figure \ref{fig5}(b), the evolution of the total energy for the forced
models, F1 (black), F2 (blue), and F3 (red) are shown.
In all the forced models the total energy rises steeply and then
saturates at final states since the dissipation rate balances the input
power.
The amplitude of the final saturated energy level decreases with
decreases in the collisional coupling constant, again showing that the
dissipation rate increases as the strength of ambipolar diffusion increases.

\section{Summary}

In this paper we describe specific numerical methods for incorporating
ambipolar diffusion into a multidimensional MHD code based on the total
variation diminishing scheme.
We assume the strong coupling approximation, that magnetic force and the
neutral-ion drag force in weakly ionized plasmas are almost equal and so
the plasma can be treated as a single fluid.
Since our numerical methods described in this paper are fully explicit
and maintain a second-order accuracy, it is straightforward to extend
them to parallelized versions and other geometries.
The divergence-free constraint on the magnetic field has been exactly
enforced through the flux-interpolated constraint transport scheme at
all times.
By using the super time stepping method to accelerate the timestep for
ambipolar diffusion, we remove the severe restriction on the stable
timestep that would arise at high numerical resolution and/or in strong
ambipolar diffusion.

Ambipolar diffusion has been tested through the direct comparison with
analytic solutions of diffusion problems.
We have computed test problems that include oblique C-type shocks and
the decay of Alfv\'en waves.
For both of these test problems, comparisons of numerical results to
analytic solutions are possible and they demonstrate the good accuracy
and robustness of our methods.
We have also performed simulations of the compressible MHD turbulence in
the presence of ambipolar diffusion and they confirm the ability of our
code to follow complex MHD flows.
We have shown that the dissipation rate of MHD turbulence is strongly
affected by the strength of ambipolar diffusion in both decaying
turbulence and forced turbulence.
 
This multidimensional MHD code incorporating an explicit scheme for
solving the ambipolar diffusion term allows us to study astrophysical
systems such as molecular cloud cores and protostellar discs in which
ambipolar diffusion is thought to be important.
Currently this code is being used to study the evolution of compressible
MHD turbulence with ambipolar diffusion, and results of
three-dimensional, high resolution MHD simulations will be reported
elsewhere.

\acknowledgments

This work utilized a high performance cluster at the Korea Astronomy and
Space Science Institute (KASI).
EC was supported by the KASI Postdoctoral Fellowship.
JK was supported by the Korea Science and Engineering Foundation through
the Astrophysical Research Center for the Structure and Evolution of
Cosmos and by the Korea Foundation for International Cooperation of
Science and Technology through grant K20702020016-07E0200-01610.

\clearpage

\clearpage

\begin{figure}
\begin{center}
\includegraphics[scale=0.8]{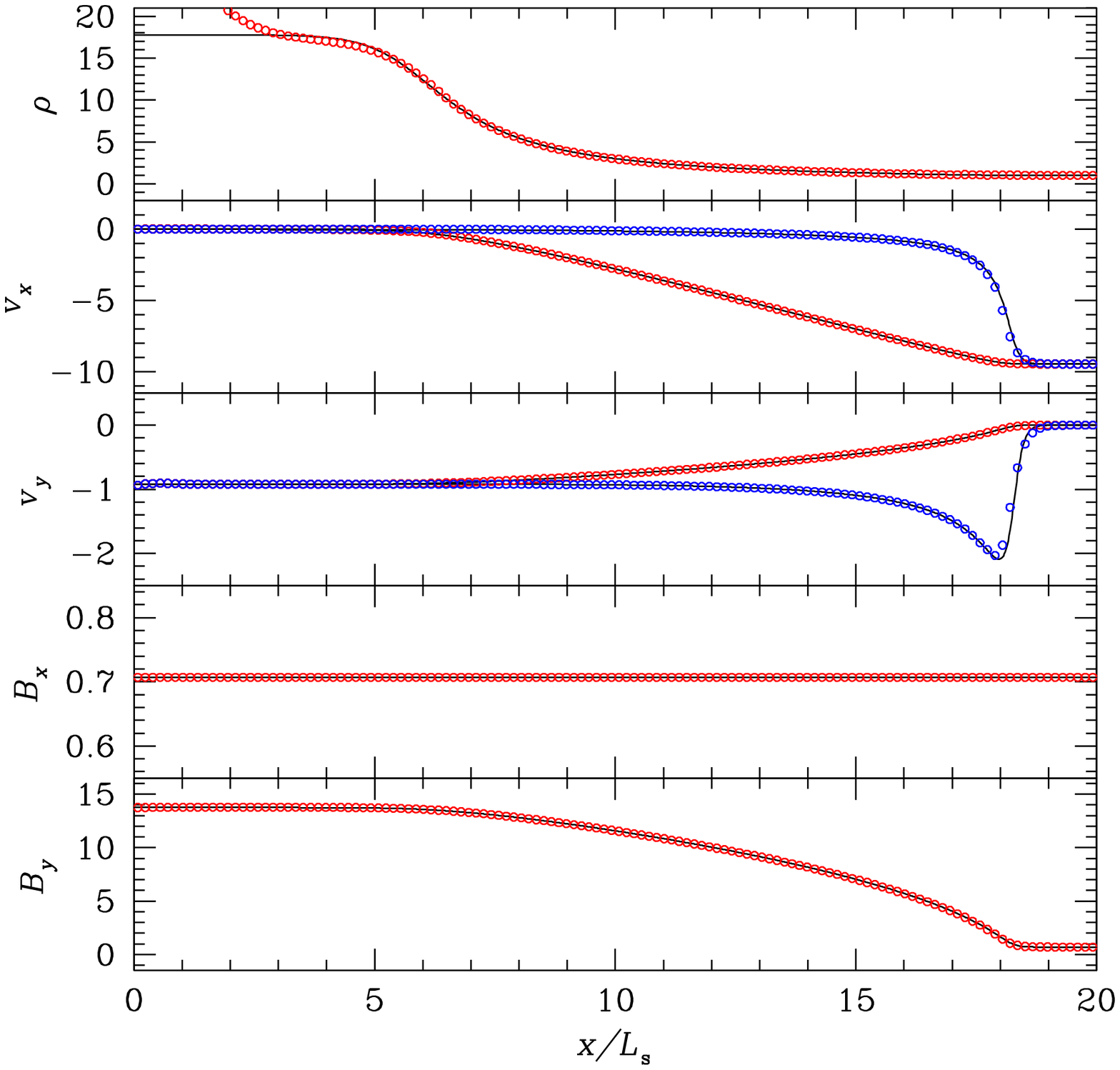}
\end{center}
%\caption{Oblique C-type shocks.}
%\label{fig1}
\end{figure}

\clearpage

\begin{figure}
\begin{center}
\includegraphics[scale=0.8]{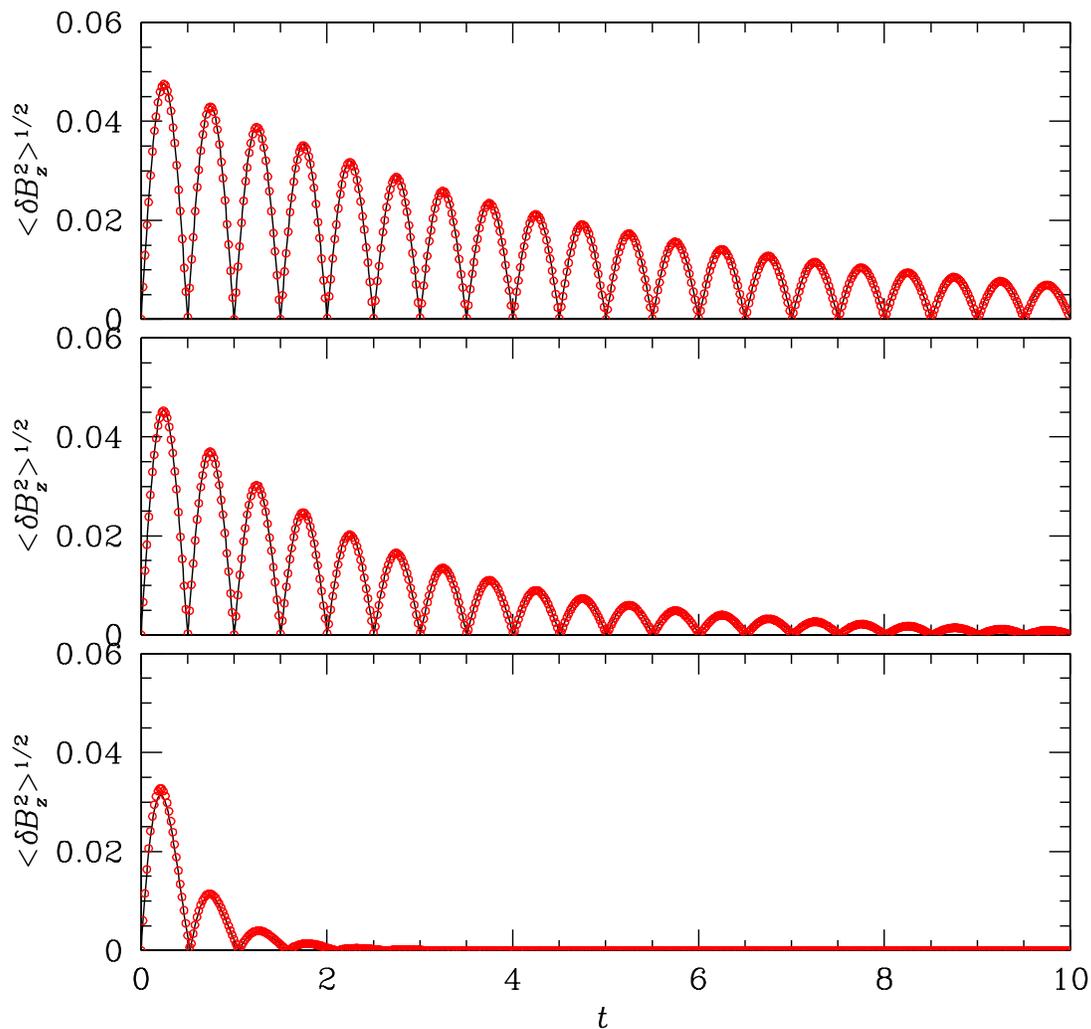}
\end{center}
\caption{Time evolution of the spatially averaged root mean square
magnetic field in the $z$-direction for $\gamma=1000$ (top), $500$
(middle), and $100$ (bottom) in the test of the decay of Alfv\'en waves.
Our numerical results (open circles) are compared to the theoretical
predictions (solid lines).
The calculations have been done in a cube box with size $L=1$ using
$128^3$ cells.
Time is expressed in units of the sound wave crossing time, $L/a$.}
\label{fig2}
\end{figure}

\clearpage

\begin{figure}
\begin{center}
\includegraphics[scale=0.8]{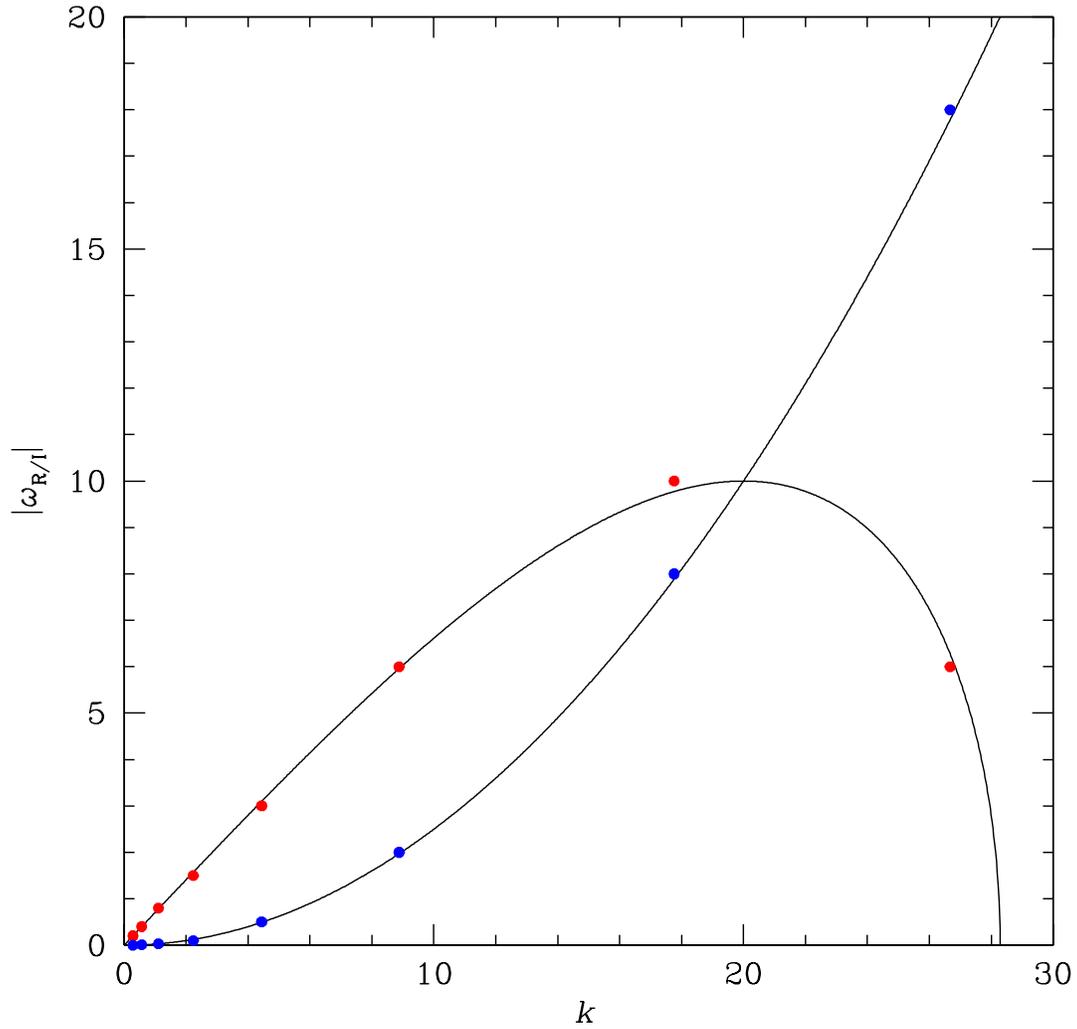}
\end{center}
\caption{Oscillation frequencies $\omega_R$ and decay rates $\omega_I$
collected from the oscillation of standing waves with different
wavenumbers for $\gamma=100$ in the test of the decay of Alfv\'en waves.
Oscillation frequencies (red) and decay rates (blue) are marked with
filled circles and the analytic solutions are drawn as solid lines.
The calculations were done in a cubic box of size $L=1$ using $128^3$
cells.}
\label{fig3}
\end{figure}

\clearpage

\begin{figure}
\begin{center}
\includegraphics[scale=0.8]{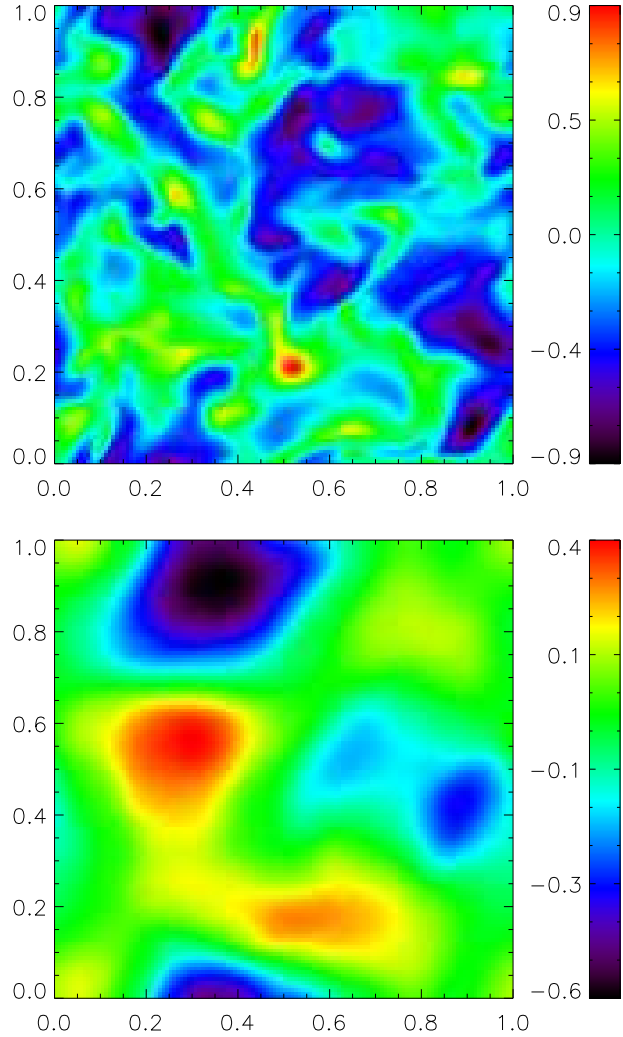}
\end{center}
\caption{(a) Images of the neutral density for
models D1 (top) and D3 (bottom) at $t=1t_f$ in the MHD turbulence
simulations.
The simulations have been done in a cubic, periodic box with size $L=1$
using $128^3$ cells, and the images shown are slices through the $x-y$
plane at $z=0.5$.
The color bars are drawn in logarithmic (base 10) scales.
(b) Same as in (a) except for models F1 (top) and F3 (bottom) at $t=1t_f$.}
\label{fig4}
\end{figure}

\clearpage

\begin{figure}
\begin{center}
\includegraphics[scale=0.8]{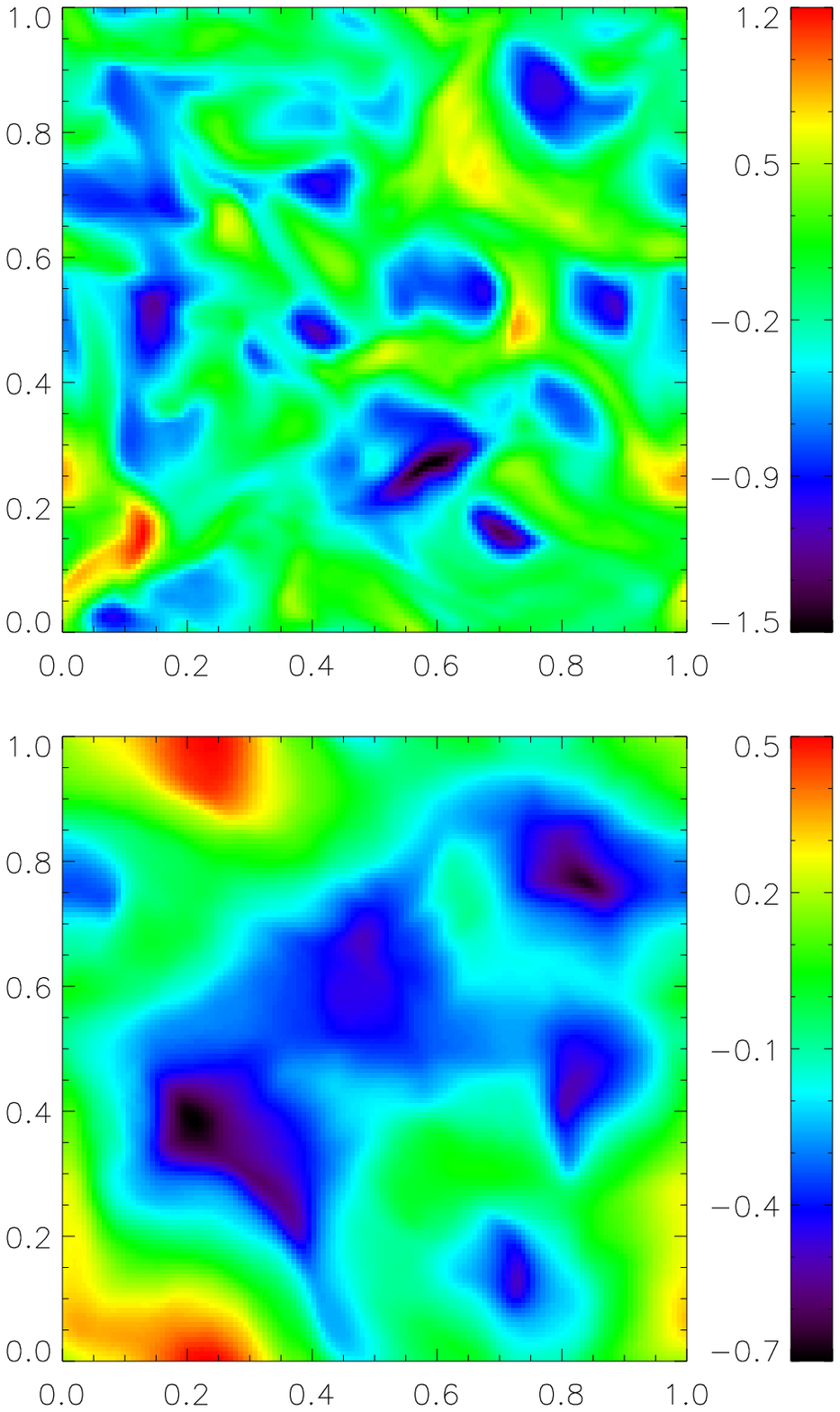}
\end{center}
%\caption{MHD turbulence simulations.}
%\label{fig4}
\end{figure}

\clearpage

\begin{figure}
\begin{center}
\includegraphics[scale=0.8]{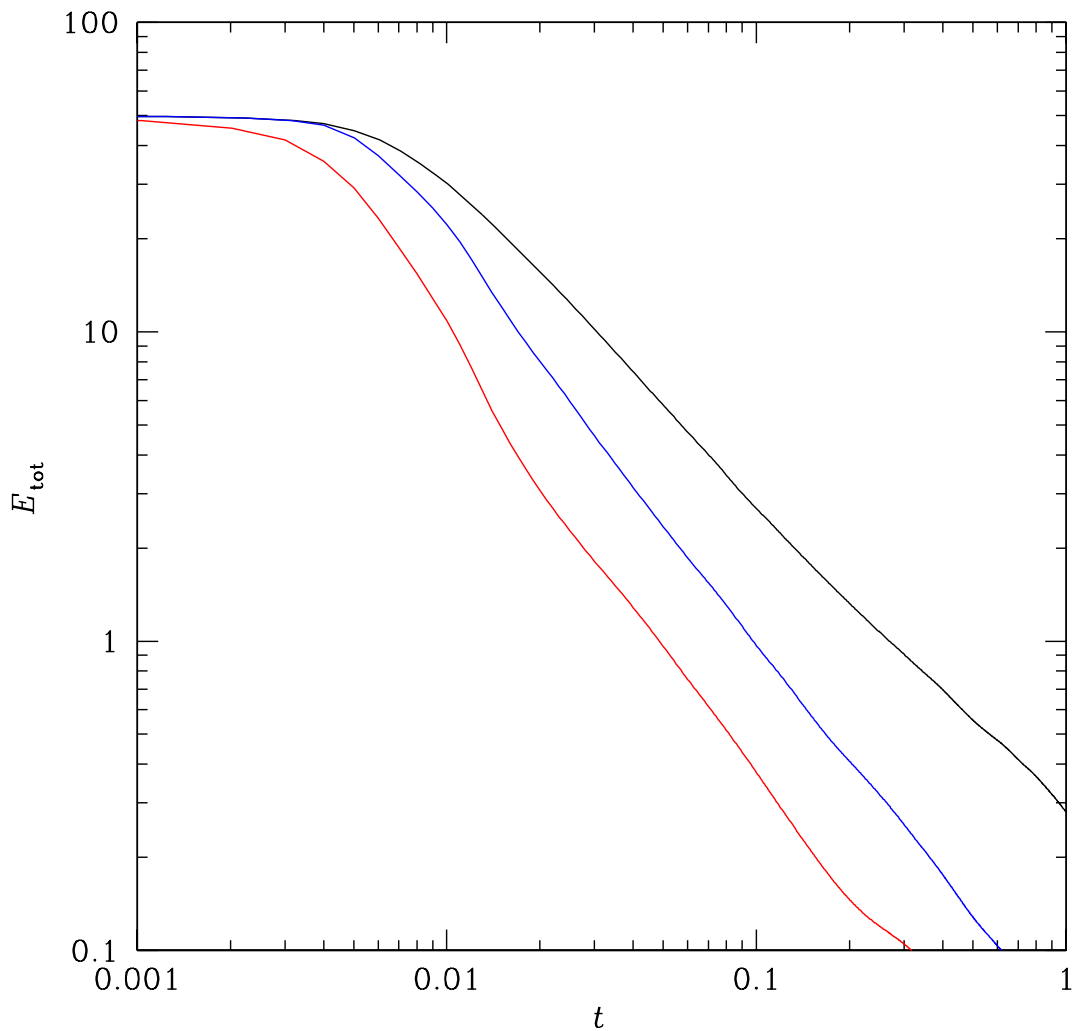}
\end{center}
\caption{(a) Time evolution of the total energy for the decaying models D1
(black), D2 (blue), and D3 (red) in the MHD turbulence simulations.
The simulations have been done in a cubic, periodic box with size $L=1$
using $128^3$ cells.
Time is expressed in units of the sound wave crossing time, $L/a$.
(b) Same as in (a) except for the forced models F1 (black), F2 (blue), and
F3 (red).}
\label{fig5}
\end{figure}

\clearpage

\begin{figure}
\begin{center}
\includegraphics[scale=0.8]{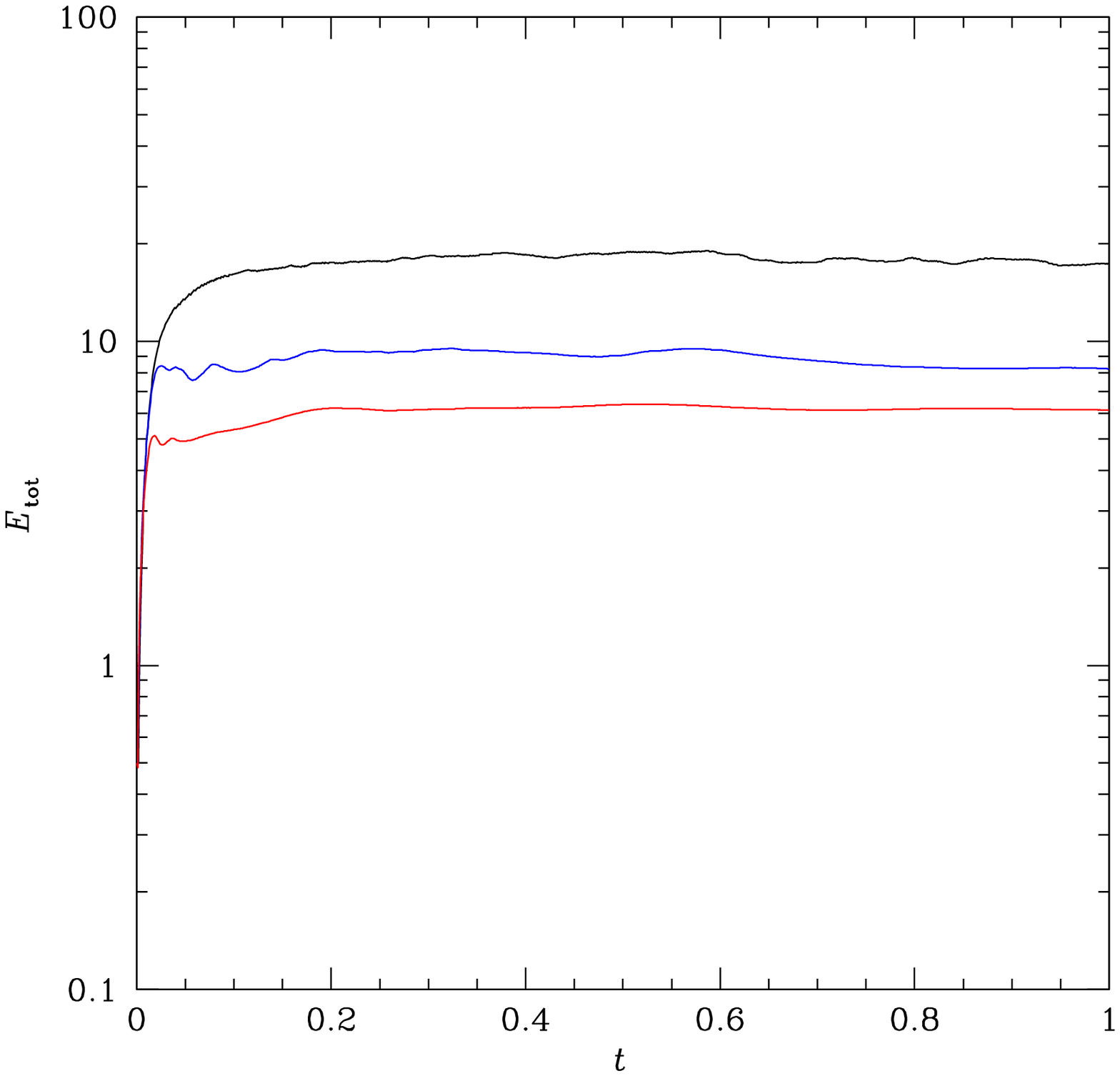}
\end{center}
%\caption{MHD turbulence simulations.}
%\label{fig5}
\end{figure}

\clearpage

\begin{deluxetable}{ccccccccc}
\tablewidth{0pt}
\tablecaption{Parameters for MHD Turbulence Simulations
\label{tab1}}
\tablehead{
\colhead{Model} &
\colhead{$\beta$} &
\colhead{$\gamma$} &
\colhead{$M_\mathrm{A}$} &
\colhead{$\tau$} &
\colhead{$R_\mathrm{AD}$} &
\colhead{$l_\mathrm{AD}$} &
\colhead{$t_f$} &
\colhead{$t_\mathrm{end}$}}
\startdata
 D1(F1) & 1 & $\infty$ & 10(1) & 0   & $\infty$ & 0         & 0.1(1) & 10(1)$t_f$ \\
 D2(F2) & 1 & 1000     & 10(1) & 0.1 & 100(10)  & 0.01(0.1) & 0.1(1) & 10(1)$t_f$ \\
 D3(F3) & 1 & 100      & 10(1) & 1   & 10(1)    & 0.1(1)    & 0.1(1) & 10(1)$t_f$ \\
\enddata
\tablecomments{Here models D1 to D3 represent decaying turbulence and
models F1 to F3 denote forced turbulence.
All the decaying and forced models have been done in pairs, and the
different values for both models are distinguished using parentheses.}
\end{deluxetable}

\end{document}